\documentclass[prd, twocolumn,letterpaper, amsmath, amssymb, preprintnumbers, showpacs, floatfix, nofootinbib, superscriptaddress]{revtex4-1}
\usepackage{mystyle_emacs}
\usepackage{theoremref}
\usepackage{placeins}
\usepackage{multirow}
\usepackage{tikz}
\usepackage{graphicx}
\usepackage{xcolor}
\usepackage[labelformat=empty]{subcaption}

\newcommand\newsubcap[1]{\phantomcaption%
       \caption*{\textbf{#1}}}

\newcommand\capa{\textbf{(a)}\,}
\newcommand\capb{\textbf{(b)}\,}

\makeatletter
\let\sv@endpart\@endpart
\def\@endpart{\thispagestyle{empty}\sv@endpart}
\makeatother
\def\xr{3}
\def\yr{3}
\usetikzlibrary{decorations.markings,decorations.pathmorphing}

\renewcommand{\ip}[2]{\left(#1,#2\right)}

\newcommand{\bco}{\bar{\mathcal{O}}}
\begin{document}

\title{Large-time correlation functions in bosonic lattice field theories}
\author{Cagin Yunus}
\email{cyunus@mit.edu}
\affiliation{Center for Theoretical Physics,
Massachusetts Institute of Technology, Cambridge, MA 02139, USA}
\author{William Detmold}
\email{wdetmold@mit.edu}
\affiliation{Center for Theoretical Physics,
Massachusetts Institute of Technology, Cambridge, MA 02139, USA}
\affiliation{The NSF Institute for Artificial Intelligence and Fundamental Interactions}

\begin{abstract}

Large-time correlation functions have a pivotal role in extracting particle masses from Euclidean lattice field theory calculations, however little is known about the statistical properties of these quantities. In this work, the asymptotic form of the distributions of the correlation functions at vanishing momentum is determined for bosonic interacting lattice field theories with a unique gapped vacuum. It is demonstrated that the deviations from the asymptotic form at large Euclidean times can be utilized to determine the spectrum of the theory.

\end{abstract}

\preprint{{MIT-CTP-5465}}
\maketitle

\newpage

\section{Introduction}
Quantum field theory (QFT) has an essential role in nuclear and particle physics and in condensed matter physics. 
From the Standard Model of particle physics to topological phase structures in quantum materials, the language of QFT provides a concise and predictive mathematical description. In some cases these descriptions contain a small parameter (coupling) in which an expansion can be performed to derive analytical expressions for relevant physical quantities. However in other cases, the couplings are large and numerical approaches are required to extract physical results; the dominant such approach in many strongly-coupled theories is referred to as lattice QFT (LQFT). 
The LQFT method involves stochastic Monte-Carlo sampling of the very high-dimensional lattice-regulated path integrals that define the correlation functions of the theory in which physical information is encoded.
The large Euclidean-time behaviour of correlation functions plays a crucial role in LQFT as it is central to obtaining the energy spectrum of the theory under study. 
If $\co(x)$ is a localised operator built from combinations of the fundamental fields centered around the spacetime point $x=(t,\vec{x})$ and $\overline{\co}(t) \propto \sum_{\vec{x}} \co(x)$ then, for Euclidean theories satisfying reflection positivity, the bilocal operator $C(t) \equiv \overline{\co}(t) \overline{\co}^\dagger(0)$ has (vacuum, $|\Omega\rangle$) expectation value
\bad 
\ev{C(t)} = \langle \Omega | C(t) | \Omega\rangle \stackrel{t\to\infty}{\longrightarrow} Z_C e^{-mt},
\ead 
where $m$ is the mass of the lowest energy eigenstate that $\overline\co^\dagger(0)\ket{\Omega}$ has an overlap with and $Z_C$ is the overlap factor. Therefore, for large enough\footnote{``Large'' implies that $t^{-1}$ is much smaller than the difference between the masses of the ground state and the first excited state with zero spatial momentum.} $t$, an estimator for $m$ is given by
\begin{equation}
\hat m = \ln{\ev{C(t)}}-\ln{\ev{C(t+1)}}.
\end{equation}
In practice, to extract $m$, one calculates the sample mean and sample variance of $\hat m$ and assumes the validity of the Central Limit Theorem to construct confidence intervals. However, for most choices of correlation function, as the Parisi-Lepage \cite{Parisi:1983ae,Lepage:1989hd} argument shows, the signal-to-noise ratio vanishes exponentially fast as $t \to \ii$.\footnote{A notable exception is the correlation function built from local pseudoscalar interpolating operators.} This issue has been investigated in detail in lattice calculations in the context of lattice Quantum Chromodynamics (LQCD), given its phenomenological interest as the theory of the strong interaction. Recent works have focused on characterising the nature of statistical fluctuations \cite{Beane:2009gs,Endres:2011jm,Endres:2011mm,DeGrand:2012ik,Grabowska:2012ik,Nicholson:2012xt,Drut:2015uua,Wagman:2016bam} and on proposing strategies for noise reduction \cite{DellaMorte:2007zz,DellaMorte:2008jd,DellaMorte:2010yp,Detmold:2014hla,Majumdar:2014cqa,Ce:2016ajy,Ce:2016idq,Wagman:2017xfh,Wagman:2017gqi,Detmold:2018eqd,Porter:2016vry,DallaBrida:2020cik,Detmold:2020ncp,Kanwar:2021wzm,Detmold:2021ulb}.

In confronting noise issues in the large-time behaviour of correlation functions, it would be useful to have an analytical form for the distribution of $C(t)$ even in simple examples. Such a form would allow for statistical tests of empirical distributions determined in numerical calculations that may diagnose statistical limitations in the empirical sampling. In the present work, it will be demonstrated that in certain cases it is possible to obtain such analytical forms in the large time regime for correlation functions whose  source and sink are both constructed to have vanishing spatial momentum. In Sec.\ \ref{sec:free_scalar}, the analytical form of correlation function distributions for the free real scalar field theory will be derived. In Sec.\ \ref{sec:long_time}, it will be shown that the distribution obtained in Sec.\ \ref{sec:free_scalar} is valid for correlation functions constructed from local operators for interacting bosonic theories at large times assuming a unique, gapped vacuum. Deviations from the asymptotic form of the distribution are linked to excited states in the spectrum and are controlled by the correlation lengths (or masses) of the theory. Consequently, analysis of these deviations provides a means of determining  correlation lengths.
Through their more general approach to statistical distributions, these results can potentially provide a path towards more robust determinations of energies and matrix elements in LQFT.

\section{Exact Results for Free Scalar Field Theory}
\label{sec:free_scalar} 
In this section, the distributions of correlation functions whose source and  sink  are both constructed to have vanishing spatial momentum are studied in the setting of a free real scalar field theory. The general form of the distribution is derived before a specific lattice discretisation is chosen and investigated.

Since the free theory momentum modes decouple, all non-zero spatial momentum modes can be factored out trivially. That is, if $\phi(t)$ is defined as
\bad 
\phi(t) = \ff{1}{\ss V}\sum_{\vec x} \phi(t,\vec x),
\ead 
\bad 
Z = \oint_{\phi(0)=\phi(N_t)} [{\cal D}\phi] \exp{-\ff 1 2 \sum_{t',t''}\phi(t')D(t',t'')\phi(t'')},
\label{eq:Zfree}
\ead 
where $N_t$ is the lattice size in the temporal direction, $V$ is the spatial volume in lattice units,  $D(t',t'')$ is a discretisation of the continuum Klein-Gordon operator, and $[{\cal D}\phi]=\prod_{t=0}^{N_t-1} d\phi(t)$ is the integration measure.

For the following argument, it is assumed that the discretised $D(t',t'')$ is 
\begin{itemize}
\item positive-definite, in order that the integral in Eq. \eqref{eq:Zfree} is convergent,
\item real and symmetric, 
\item translationally invariant, only depending on $t'-t'' (\text{mod } N_t)$.
\end{itemize}

The characteristic function for the composite operator $\phi(t) \phi(0)$ is defined as 
\begin{widetext}
\bad 
\Phi_{\phi(t) \phi(0)}(\o) &= \ev{\exp{-i\o \phi(t) \phi(0)}} \\ 
&= \ff{1}{Z} \oint_{\phi(0)=\phi(\b)} \lk {\cal D}\phi \rk \exp{-\ff 1 2 \sum_{t',t''}\phi(t')D(t',t'')\phi(t'')-i\o \phi(t)\phi(0)}.
\ead 
\end{widetext}
If $Q_{t',t''} = \d_{t',0}\d_{t'',t} + \d_{t',t}\d_{t'',0}$ is further defined, then it is easy to show that
\bad 
\Phi_{\phi(t) \phi(0)}(\o) &= \ff{1}{\ss{\det R(\o)}},
\ead 
where 
\be 
R(\o) = 1 + i\o D^{-\ff 1 2}QD^{-\ff 1 2}.
\label{eq:defR}
\ee 

Note that $D$ and $Q$ are linear operators acting on a vector space of dimension $N_t$ and let $e_{t'}$ for $t'=0,\cds,N_t-1$ be the basis of unit vectors on each timeslice and for which matrix elements are given as $D_{t',t''}$ and $Q_{t',t''}$ respectively. 
Note that in this basis $Qe_0 = e_t$ and $Qe_t = e_0$ and $Qe_{t'} = 0$ for $t' \not \in \{0,t\}$. Let $(\cd,\cd)$ be the inner product for which the vectors $e_{t'}$ are orthonormal. For the basis $\{v_{t'} = D^{\ff 1 2}e_{t'}\}$, $R(\o)v_{t'} = v_{t'}$ for $t' \not \in \{0,t\}$. Therefore, to calculate $\det R(\o)$, only the subspace spanned by $v_0$ and $v_t$ needs to be considered. Further, the vectors $v_\pm = D^{\ff 1 2} e_\pm$ can be defined where $e_\pm = \ff{1}{\ss 2}\lp e_t \pm e_0 \rp$. If $R(\o)v_+ = R_{++}v_+ + R_{+-}v_- + \cds $ and $R(\o)v_- = R_{--}v_- + R_{+-}v_+ + \cds $, where the ellipsis represents terms that are spanned by $v_{j}$ for $j \neq 0,t$ and does not contribute to the determinant, then:
\be 
\det R(\o) = R_{++}R_{--}-R_{+-}R_{-+}.
\ee
Further, for  $w_{t'} = D^{-\ff 1 2}e_{t'}$ and $w_\pm = D^{-\ff 1 2}e_\pm$. Then, $(w_\pm,v_\pm) = \d_{\pm,\pm}$. It follows that 
\bad 
R_{\s\s'} &= \d_{\s,\s'} + i\o \s' \ip{e_{\s}}{D^{-1}e_{\s'}},
\label{eq:Rss'}
\ead
where $\s,\s' = \pm$. From translational invariance, it also follows that 
$\ip{e_{\pm}}{D^{-1}e_{\mp}} = \ip{e_{0}}{D^{-1}e_{0}} - \ip{e_{t}}{D^{-1}e_{t}} \pm \ip{e_{t}}{D^{-1}e_{0}} - \ip{e_{0}}{D^{-1}e_{t}} =0$ and 
so $R_{+-}=R_{-+}=0$. Then, by defining\footnote{The positivity of $\o_\pm$ follows from the positive-definiteness of $D$.}
\bad 
\o_{\pm} = \ff{1}{\ip{e_{\pm}}{D^{-1}e_{\pm}}}>0,
\label{eq:defopm}
\ead 
it follows that
\bad 
\det(R(\o)) = \ff{\lp\o-i\o_+\rp\lp\o+i\o_-\rp }{\o_+ \o_-}.
\ead 
Therefore, the probability distribution of $\phi(t)\phi(0)$ is obtained from the characteristic function as:
\bad 
P_{\phi(t)\phi(0)}(x) &\equiv \ff{1}{2\pi}\int d\o\, e^{i\o x}\Phi_{\phi(t)\phi(0)}(\o)\\
&=\ff{\ss{\o_+\o_-}}{2\pi}\int d\o\, \ff{e^{i\o x}}{\ss{(\o-i\o_+)(\o+i\o_-)}}.
\label{eq:FreeProductPIntegral}
\ead 

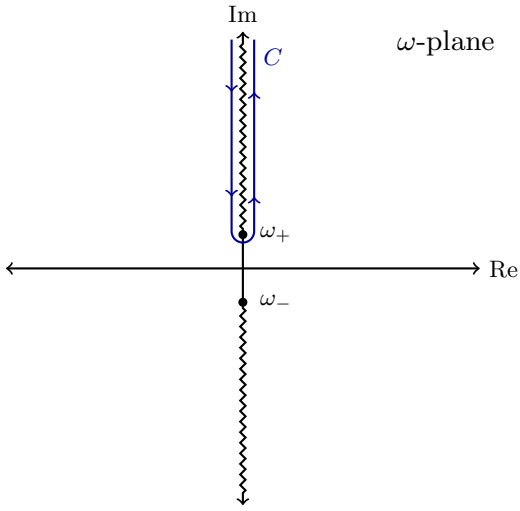
\begin{figure}
\centering
\begin{tikzpicture}[thick]
  \node[scale=1.2] at (2.7,3.0) 
    {$\o$-plane};
  \draw [->] (0,0) -- (-1.05*\xr,0) ;
  \draw [->] (0,0) -- (1.05*\xr,0) node [right] {Re};
  \draw [->,decorate,decoration={zigzag,segment length=4,amplitude=1,pre=lineto,pre length=15,post=lineto,post length=3}] (0,0) -- (0,-1.05*\yr) ;
  \draw [->,decorate,decoration={zigzag,segment length=4,amplitude=1,pre=lineto,pre length=15,post=lineto,post length=3}] (0,0) -- (0,1.05*\yr) node [above] {Im};
  \draw[yshift=9.8,blue!60!black,decoration={markings,mark=between positions 0.125 and 0.875 step 0.25 with \arrow{>}},postaction={decorate}] (-\xr/20,9*\yr/10) -- (-\xr/20,\yr/20) arc (-180:0:\xr/20) (\xr/20,\xr/20) -- (\xr/20,9*\yr/10) node[below right] {$C$};
  \filldraw [black] (0,0.45) circle (1.3pt)  node [right] {$\ \omega_+$};
  \filldraw [black] (0,-0.45) circle (1.3pt)  node [right] {$\ \omega_-$};
\end{tikzpicture}
\caption{The integration contour used to evaluate I(x) for $x>0$.}
\label{fig:contour}
\end{figure}
\FloatBarrier

The calculation of $P_{\phi(t)\phi(0)}$ then reduces to the evaluation of 
\begin{equation}
I(x) = \int_{-\ii}^{\ii} d\o\, \ff{e^{i\o x}}{\ss{\lp \o - i\o_+ \rp \lp \o + i\o_-\rp}}.
\end{equation}
Since $\o_\pm > 0$ (Eq.~\eqref{eq:defopm}),  for $x>0$ the integration contour in $I(x)$ can be deformed to the contour $C\equiv\{i \ii -\e \to i\o_+ -\e \} \cup \{i\o_+ + \e \to i\ii+\e\}$ as in Fig. \ref{fig:contour}.
The integral can thus be expressed as: 
\be 
 I(x) = \lp -e^{i\ff{3\pi}{4}} + e^{-i\ff{\pi}{4}}\rp ie^{-i\ff{\pi}{4}} I_0(x) = 2\bar{I}(x),
\ee 
where 
\bad
\bar{I}(x) &= \int_{\o_+}^{\ii} dy\ff{e^{-x y}}{\ss{\lp y-\o_+ \rp \lp y+\o_- \rp}}\\
&= e^{\ff 1 2 \lp \o_- - \o_+ \rp x}K_0 \lp \ff{1}{2}\lp \o_+ + \o_- \rp x \rp .
\ead  
Here, $K_0(x)$ is a modified Bessel function of the second kind. These formulas can be summarised as\footnote{ Note that the similarity of this equation to the form of the scalar field propagator in position space, $$C(x)\sim \frac{m^{(d-1) / 2}}{|x|^{(d-1) / 2}} K_{(d-1) / 2}\left(m |x| \right)
	\stackrel{d\to1}{\longrightarrow}K_{0}\left(m |x|\right),$$ 
	is coincidental for $d=1$ and does not hold in other numbers of spatial dimensions, $d$.}:
\be 
P_{\phi(t)\phi(0)}(x) = \ff{\ss{\o_+ \o_-}}{\pi}e^{\ff 1 2 \lp \o_- - \o_+\rp x}K_0 \lp \ff{1}{2}\lp \o_+ + \o_- \rp \abs{x} \rp. 
\label{eq:FreeProductP}
\ee 

The lattice action needs to be specified explicitly to gain more insight into $\o_\pm$. A simple discretisation is:
\be 
\lp D\cd \phi \rp(t) = m^2 \phi(t) - \lp \phi(t+1)+\phi(t-1)-2\phi(t)\rp.
\label{eq:Ddef}
\ee
Assuming that $N_t$ is odd, the normalised eigenvectors of $D$ are given by: 
\bad 
E(t) &= \ff{1}{\ss N_t},\\ S_j(t) &= \ss{\ff{2}{N_t}} \sin{\ff{2\pi j}{N_t}t},\\ C_j(t) &= \ss{\ff{2}{N_t}}\cos{\ff{2\pi j}{N_t}t},
\label{eq:Deigvecs}
\ead
where $j=1,\cds,\ff{N_t-1}{2}$ and $t$ indexes the components of the each vector. The corresponding eigenvalues are: 
\bad \l(E)&=m^2, \\ 
\l(S_j) &=  m^2+4\sin{\ff{\pi j}{N_t}}^2,   \\
\l(C_j) &=  m^2+4\sin{\ff{\pi j}{N_t}}^2. 
\label{eq:Deigvals}
\ead 

Expanding in eigenvectors of $D$, the vector representing an arbitrary zero momentum field $\phi$ can be expressed as $\phi(t) = \phi^e E(t) + \sum_{j=1}^{\ff{N_t-1}{2}} \phi^s_j S_j(t) + \sum_{j=1}^{\ff{N_t-1}{2}} \phi^c_j C_j(t)$ and it follows that
\bad 
\lc \phi(t) \phi(0) \rc &= \ff{1}{N_t}\lc (\phi^e)^2 \rc + \ss{\ff{2}{N_t}}\sum_{j=1}^{\ff{N_t}{2}} C_j(t)\lc (\phi^c_j)^2 \rc \\ 
&= \ff{1}{N_t}\ff{1}{m^2} + \ff{2}{N_t} \sum_{j=1}^{\ff{N_t-1}{2}} \ff{\cos{\ff{2\pi j}{N_t}t}}{m^2 + 4\sin{\ff{\pi j}{N_t}}^2},
\ead
where $\langle \ldots \rangle$ indicates integration over the field $\phi$ where the integration measure is given by $\lk \cald \phi \rk = \prod_{t=0}^{N_t-1} d\phi(t) = d\phi^e \prod_{j=1}^{\ff{N_t-1}{2}} d\phi^s_j d\phi^c_j$.

Partially following Ref.~\cite{smit_2002}, in order to take the continuum limit, the infinitesimal time interval $\epsilon$ is introduced and the limit $t,N_t\to \ii$ is considered such that $\beta= \epsilon N_t$ and $t/N_t$ are fixed. Further, $\tt \equiv t\epsilon$ is defined.  The quantity $m$  depends on $\epsilon$ (equivalently on $N_t$) and $m(\epsilon)$ should be chosen such that the correlation function decays as $~\exp{-m_R
	\tt}$ as $\epsilon \to 0 $ for large $\tt$ and $\beta\to \ii$ (the $\beta\to \ii$ limit must be taken first), where $m_R$ is the renormalized mass. The renormalized field $\phi_R = \ss{Z_\phi(\epsilon)}\phi$, where $Z_\phi(\epsilon)$ will be chosen such that correlation functions of $\phi_R$ have non-singular behaviour as the continuum limit is taken. Setting $k = \ff{2\pi}{\beta}j$, the renormalized correlation function is: 
\begin{widetext}
\bad 
\lc \phi_R(\tt)\phi_R(0) \rc  &\equiv Z_\phi(\epsilon) \lc \phi(t) \phi(0) \rc \\ 
&=\ff{Z_\phi(\epsilon)\epsilon}{\pi} \sum_{k=\ff{2\pi}{\b}}^{\ff{\pi}{\b}(N_t-1)}  \ff{\D k \cos{k\tt}}{m(\epsilon)^2 + 4 \sin{\ff{k}{2}\epsilon}^2}  + \ff{Z_\phi(\epsilon)}{N_tm(\epsilon)^2}  \\ 
&= -\ff{Z_\phi(\e)\e}{\b m(\e)^2} + \ff{Z_\phi(\epsilon)\epsilon}{\pi} \sum_{k=0}^{\ff{\pi}{\b}(N_t-1)}  \ff{\D k \cos{k\tt}}{m(\epsilon)^2 + 4 \sin{\ff{k}{2}\epsilon}^2}.
\ead
\end{widetext}
Taking the limit $\b\to \ii$ (or equivalently $N_t \to \ii$) one therefore obtains:
\bad 
\lim_{\b \to \ii} \lc \phi_R(\tt) \phi_R(0) \rc &= \ff{Z_\phi(\epsilon)\epsilon}{\pi}\int_{0}^\ff{\pi}{\epsilon} dk\ff{\cos{k\tt}}{m(\epsilon)^2 + 4 \sin{\ff{k}{2}\epsilon}^2}  \\ 
&= \ff{Z_\phi(\epsilon)\epsilon}{2\pi}\int_{-\ff{\pi}{\epsilon}}^\ff{\pi}{\epsilon} dk\ff{\exp{i k\tt}}{m(\epsilon)^2 + 4 \sin{\ff{k}{2}\epsilon}^2}.
\ead

The integration can be performed by defining $z = \exp{i k \epsilon}$ that maps onto the unit circle, giving\footnote{Note that $\ff{\tt}{\epsilon}$ is a non-negative integer, so the integrand is meromorphic.}:
\bad 
\lc \phi_R(\tt)\phi_R(0) \rc = - \ff{Z_\phi(\epsilon)}{2\pi i} \oint dz \ff{z^{\ff{\tt}{\epsilon}}}{z^2 - z(m(\epsilon)^2 + 2) + 1  }.
\ead
The poles of the integrand are at $z_\pm = 1 + \ff{m(\epsilon)^2}{2} \pm m(\epsilon)\ss{1+\ff 1 4 m(\epsilon)^2}$, with only $z_-$ occurring inside the unit circle. Therefore:
\bad 
\lc \phi_R(\tt)\phi_R(0) \rc &= \ff{Z_\phi(\epsilon)}{2m(\epsilon)\ss{1+\ff 1 4 m(\epsilon)^2}} \\&\quad\times \lp  1 + \ff{m(\epsilon)^2}{2} - m(\epsilon)\ss{1+\ff 1 4 m(\epsilon)^2} \rp^{\ff{\tt}{\epsilon}}.
\ead 
To observe the expected exponential decay $\propto \exp{-m_R \tt}$, one is forced to set $m(\epsilon) = m_R \epsilon + \co \lp \epsilon^2 \rp $. This also shows that one must chose  $Z(\epsilon) = \text{constant} \times \epsilon + \co \lp \epsilon^2 \rp$ for the above equation to be finite for $\epsilon\to0$. 
Given these constraints, $m(\e) = m_R \e$ and $Z_\phi(\e) = \e$ are chosen, leading to a renormalised correlation function that is finite in the $\epsilon\to0$ limit.

It is clear from Eqs. \eqref{eq:Deigvecs} and \eqref{eq:Deigvals} that\footnote{ Note that this is a vector equation: both $D^{-1}e_t$ and $E$, $S_j$ and $C_j$ have $N_t$ components.}
\bad 
\inv De_{t} &= \ff{1}{m_R^2\e^2\ss N_t}E \\ &+ \ss{\ff{2}{N_t}}\sum_{j=1}^{\ff{N_t-1}{2}}\ff{\sin{\ff{2\pi jt}{N_t}}S_j +\cos{\ff{2\pi jt}{N_t}}C_j}{ m_R^2\e^2+ 4\sin{\ff{\pi j}{N_t}}^2},
\ead  
so it is straightforward to show that:
\bad
\ip{e_+}{\inv D e_+} &= \ff{2}{N_t m_R^2 \e^2} + \ff{2}{N_t}\sum_{j=1}^{\ff{N_t-1}{2}} \ff{1+\cos{\ff{2\pi jt}{N_t}}}{  m_R^2\e^2+ 4\sin{\ff{\pi j}{N_t}}^2},\\
\ip{e_-}{\inv D e_-} &=  \ff{2}{N_t}\sum_{j=1}^{\ff{N_t-1}{2}} \ff{1-\cos{\ff{2\pi jt}{N_t}}}{  m_R^2\e^2+ 4\sin{\ff{\pi j}{N_t}}^2}, \\
\ip{e_+}{\inv D e_-} &= \ip{e_-}{\inv D e_+}\\
&= 0.
\ead
The probability density function of the renormalized two-point function is given as:
\be 
P_{\phi_R(\tt)\phi_R(0)}(x) = \expval{\d \lp x-\phi_R(\tau)\phi_R(0) \rp}
\ee
and since $Z(\e)=\e$ is chosen, this may be written equivalently as
\be 
P_{\phi_R(\tt)\phi_R(0)}(x) = \expval{\d \lp x- \e \phi \lp  t \rp \phi(0) \rp}.
\ee 
The factor $\e$ leads to the modification of Eqs. \eqref{eq:defR} and \eqref{eq:defopm} as 
\bad
R(\o) &= 1+i\e \o  D^{-\ff 1 2}QD^{-\ff 1 2},
\ead 
and 
\bad
\o_{\pm} &= \ff{1}{\e\ip{e_{\pm}}{D^{-1}e_{\pm}}}.
\ead 
In the limit $\e \to 0$:
\bad 
\inv \o_+ &\to \ff{2}{\beta m_R^2}+\ff{\cosh{\ff{\beta m_R}{2}}+ \cosh{\ff{\beta m_R}{2} - m_R\tt}}{2m_R \sinh{\ff{\beta m_R}{2}}},\\
\ead
and 
\bad 
\inv \o_- &\to \ff{\cosh{\ff{\beta m_R}{2}}- \cosh{\ff{\beta m_R}{2} - m_R\tt}}{2m_R \sinh{\ff{\beta m_R}{2}}},
\ead
where the following identity\footnote{The sum $f(\tt) = \sum_{k} \ff{e^{ik\tt}}{m_R^2+k^2}
$ is to be calculated where $k=\ff{2\pi}{\b}j$ with $j \in \mathbb{Z}$. One observes that $f''(\tt) = -\sum_{k}e^{ik\tt}+m_R^2f(\tt) 
$. Using the fact that $\sum_k e^{ik\tt} = \b \d(\tt)
$ under the restriction $0\leq \tt < \b$, we obtain
$f''(\tt) = -\b\d(\tt)+m_R^2f(\tt)$. Therefore, one has the solution $f(\tt) = A_+e^{-m_R\tt} + A_-e^{m_R\tt} 
$, except at $\tt=0$. The symmetries of the summation in $f(\tau)$ implies that $f(\b-\tt) = f^*(\tt) = f(\tt)$, and from this condition, one obtains $ 
A_- = e^{-m_R \b}A_+$. Finally, using the boundary condition
$f'(\b-\e)-f'(\e) = \b$  as required to satisfy the inhomogeneous differential equation for $f(\tt)$, one can also fix $ 
A_+ = \ff{\b}{4m_R}\ff{e^{\ff{\b m_R}{2}}}{\sinh{\ff{m_R\b}{2}}}$, resulting in Eq.~\eqref{eq:id}. } has been used:
\be 
\sum_{k=\ff{2\pi j}{\beta}; j\in\mathbb{Z}} \ff{e^{ik\tt}}{m_R^2+k^2} =\ff{\b}{2m_R}\ff{\cosh{\ff{\b m_R}{2} - m_R\tt}}{\sinh{\ff{\b m_R}{2}}}.
\label{eq:id}
\ee 

\begin{figure}
\includegraphics[scale=0.38]{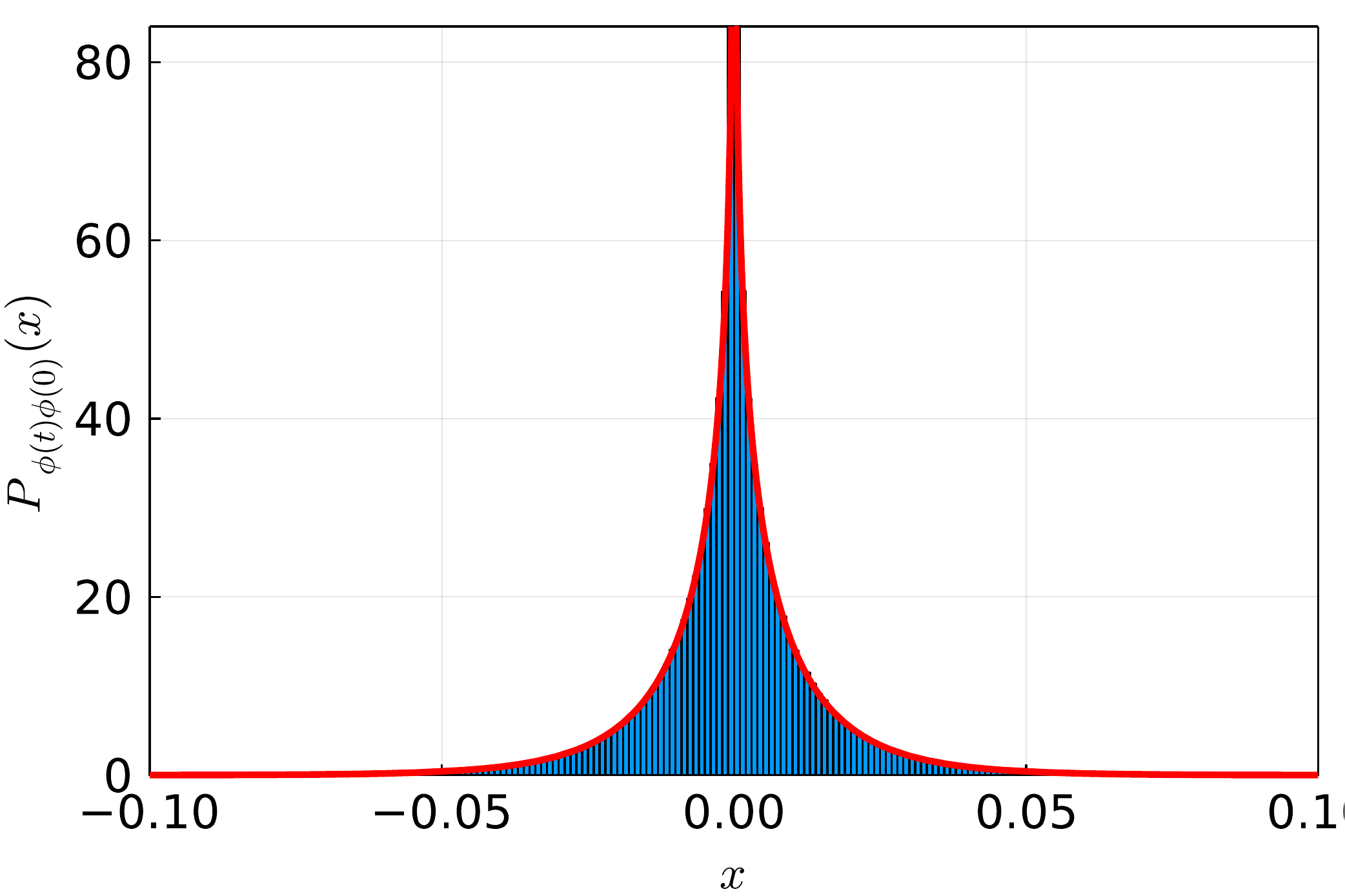}
\caption{The empirical distribution of $\phi(t) \phi(0)$ for free scalar field theory with the discretisation given by Eq. \eqref{eq:Ddef}. The red curve is given by Eq. \eqref{eq:FreeProductP} with $\o_\pm$ from Eq.~\eqref{eq:revtrans}. }
\label{fig:free}
\end{figure}

In the limit $\b \to \ii$,
\bad 
\inv \o_\pm &\to \ff{1\pm e^{-m_R \tt }}{2m_R}, 
\ead
and further in the limit $\tt \to \ii$,
\bad 
\inv \o_\pm &\to \ff{1}{2m_R}. \\ 
\ead
In the large $\tt,\b$ limit (at non-zero lattice spacing) with $\o_+ = \o_-$, Eq. \eqref{eq:FreeProductP} will be shown to valid for a much larger class of theories in the next section. First however, the distribution in Eq. \eqref{eq:FreeProductP} is compared to the numerical distribution of $\phi(t)\phi(0)$ for the two-dimensional free real scalar field theory discretised as in Eq. \eqref{eq:Ddef}. Computations are performed  for a lattice of size $N_s \times N_t =20 \times 40$ for $t=6$ and a sample size of ${\cal N}=10^6$ and the resulting distribution is shown in Fig.\ \ref{fig:free}. 

The parameters $\omega_\pm$ to use in Eq.~\eqref{eq:FreeProductP} for comparison are determined as follows. The lowest moments of the distribution in Eq. \eqref{eq:FreeProductP} are given as:
\bad 
\ev{x} &= \ff{\o_- - \o_+}{2\o_+\o_-},\\
\ev{x^2} &= \ff{3\o_-^2 - 2\o_-\o_+ + 3\o_+^2}{4\o_-^2\o_+^2},
\ead
where $\ev{x^n} = \int_{-\infty}^\infty dx\, P_{\phi(t)\phi(0)}(x)x^n$. 
After some algebra, these relations can be inverted to give
\bad 
\o_\pm &= \frac{\sqrt{\ev{x^2}-2 \ev{x}^2} \mp \ev{x}}{\ev{x^2}-3\ev{x}^2}.
\label{eq:revtrans}
\ead 

\begin{figure}[!t]
 	\begin{subfigure}{1.0\linewidth}
    	\includegraphics[width=\columnwidth]{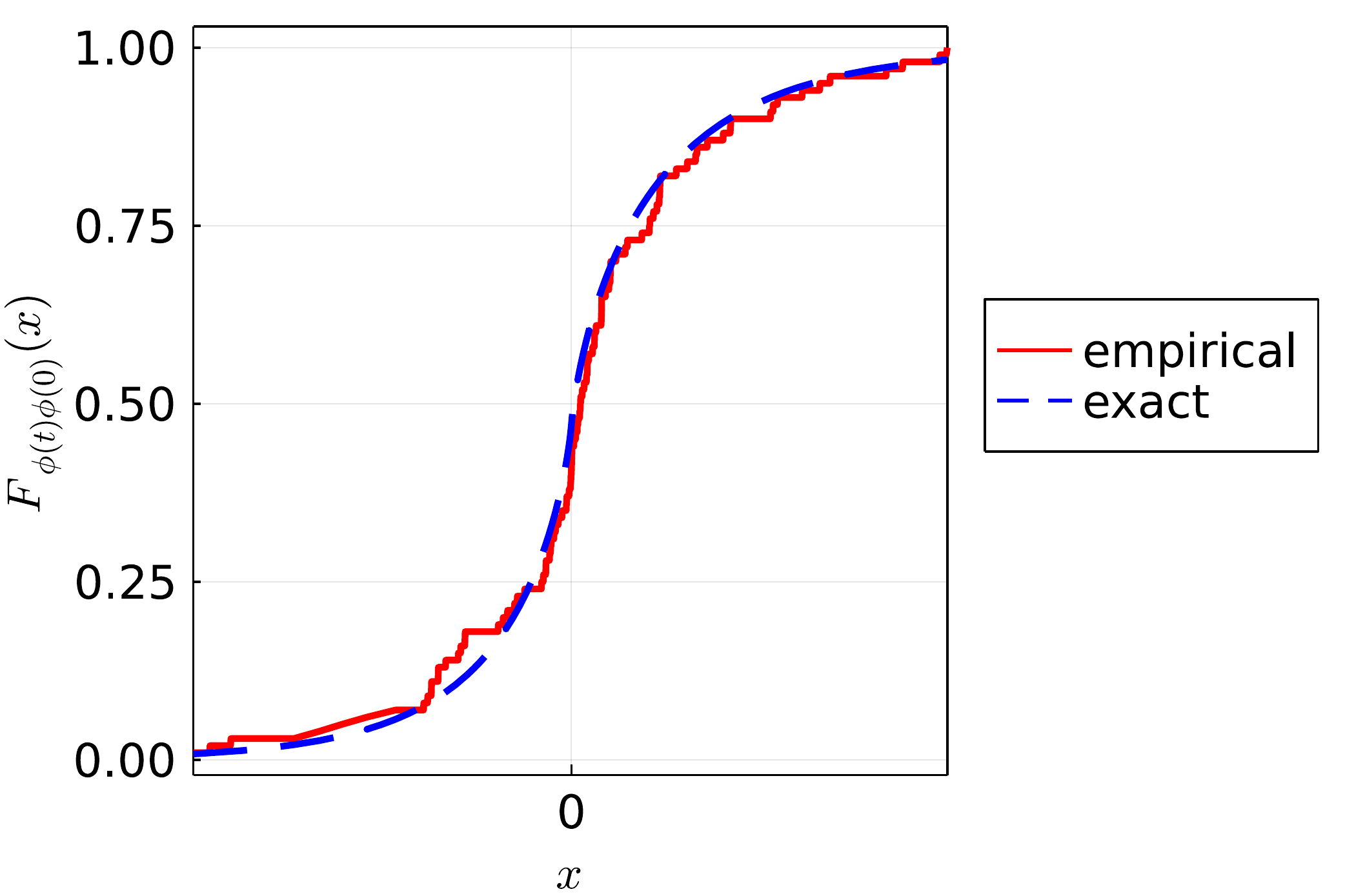}
    	\newsubcap{(a)}
	\end{subfigure}
	\begin{subfigure}{1.0\linewidth}
	\includegraphics[width=\columnwidth]{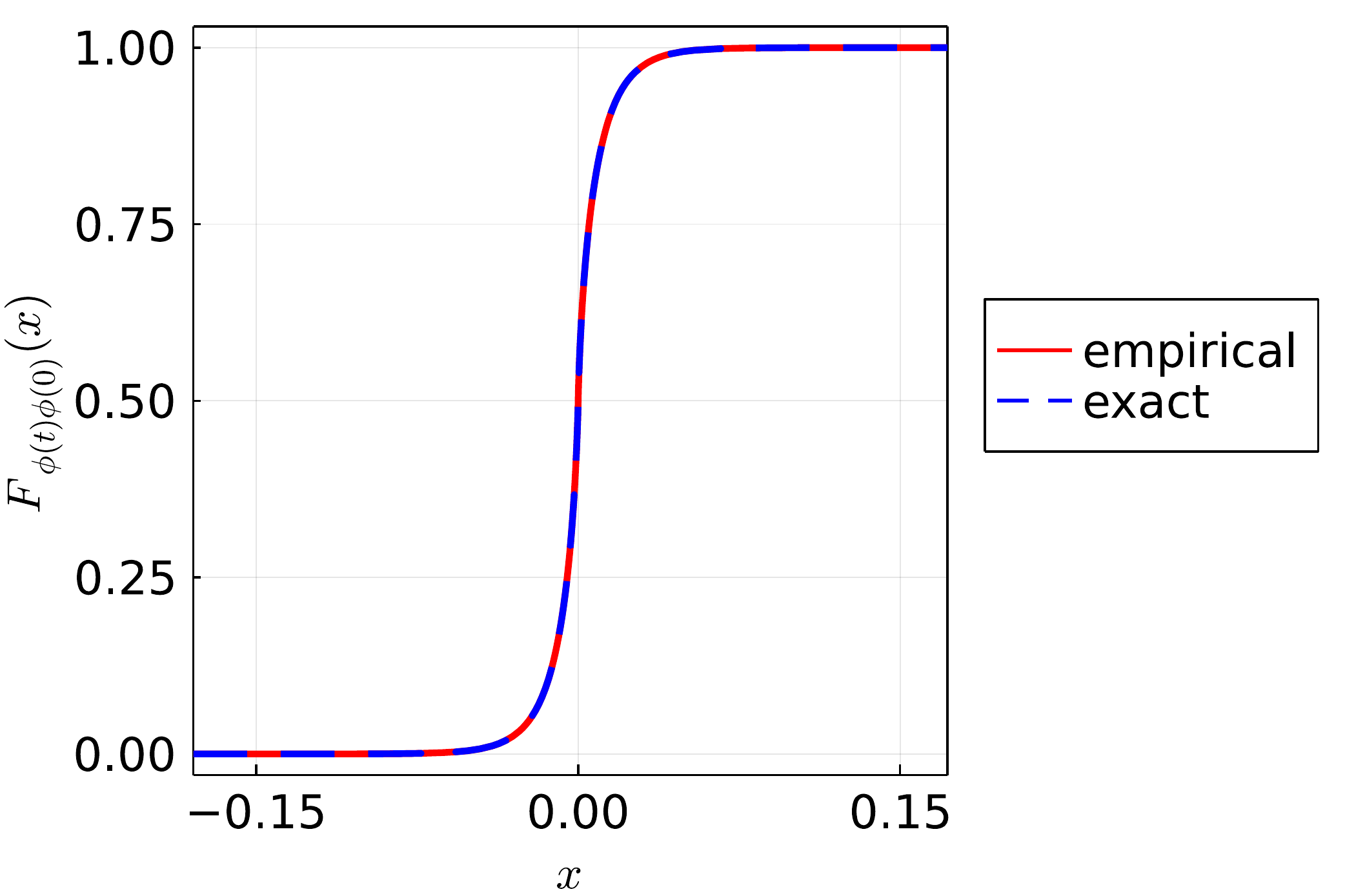}
	\newsubcap{(b)}
    \label{fig:he_3_std}
	\end{subfigure}
	\caption{\capa shows the empirical distribution function and the exact cumulative distribution function for the sample size $10^2$. \capb shows the same data and curves for sample size $10^6$.
	\label{fig:cdf}}
\end {figure}

As well as the empirical distribution from the numerical calculations, Fig.\ \ref{fig:free}  shows the distribution of Eq.~\eqref{eq:FreeProductP} with the above values of $\o_\pm$ determined from  moments of the empirical distribution.
The choice of $t=6$ and lattice size are completely arbitrary and equivalently good agreement is seen for all $t$ and for various lattice geometries.  The numerical data are clearly well-represented by the expected behaviour and the empirical cumulative distribution function  converges to the exact cumulative distribution function given by 
\be 
F_{\phi(t)\phi(0)}(x) = \int_{-\ii}^x du\, P_{\phi(t)\phi(0)}(u) 
\ee 
as the sample size is increased, as seen in Fig.\ \ref{fig:cdf} where sample sizes ${\cal N}=10^{2}$ and $10^6$ are used. 

\section{Large-Time Correlators}
\label{sec:long_time}
The above results for a free real scalar field theory can approximately describe the statistical behaviour of correlation functions for many operators at large time in a broad class of bosonic lattice field theories. Consider a theory in $D=d+1$ dimensions and a local operator\footnote{This operator does not have to be an elementary field but can be a composite operator constructed from elementary fields.} $\co(t,\vec x)$ such that: 
\begin{itemize}
\item The Euclidean action of the theory is real,
\item There is a unique, translationally invariant, gapped vacuum $\ket{\O}$,
\item $\co(t,\vec x)$ is covariant under temporal and spatial translations,
\item The vacuum expectation value of $\co(t,\vec x)$, $\expval{\co(t,\vec x)}{\O}$, vanishes.
\end{itemize}
If $\bar \co(t)$ is defined as:
\be 
\bar \co(t) = \ss{\ff{a^d}{V}}\sum_{\vec x}\co(t,\vec x),
\ee 
where $\vec x$ runs over all lattice sites on the given time slice, then
as  $t,N_t \to \ii$:
\be 
 \lim_{V\to\ii} P_{\bar \co(t)\bar \co(0)}(q) = \ff{\o}{\pi}K_0 \lp \o\abs{q}\rp + \co\lp e^{-m \min{(t,\b-t)}} \rp ,
 \label{eq:Puvexpansion}
\ee 
with $\o > 0$. Here, $|m\rangle$ is  the zero-momentum eigenstate with the smallest energy such that 
\be 
\mel{\O}{\d(\bco - u)}{m} \neq 0 \quad \text{ for some } u \in \mathbb{R}
\label{eq:gapstate}
\ee
and $m$ is its energy.

To see this, the joint probability distribution  $P_{\co(t),\co(0)}(u,v)$ that $\co(t)$ takes value $u$ and $\co(0)$ takes value $v$  will be considered. This is defined as
\begin{widetext}
\bad
P_{\bco(t),\bco(0)}(u,v) &= \oint_{\phi(N_t)=\phi(0)} [\cald \phi] e^{-S[\phi]} \d \lp \bco(t) - u \rp \d \lp \bco(0) - v \rp \\
&= \Tr\lp e^{-(N_t-t)H}\d(\bco(t)-u)e^{-tH}\d(\bco(0)-v) \rp \\
&\stackrel{ t,N_t \to \ii}{\longrightarrow} \expval{\d(\bco(0)-u)}{\O}\expval{\d(\bco(0)-v)}{\O}  + \co\lp e^{-m \min{(t,\b-t)}} \rp, 
\label{eq:Pxexpansion}
\ead
\end{widetext}
where $\phi$ is the set of elementary fields in the theory and $[\cald \phi]e^{-S[\phi]}$ is interpreted as a probability measure. The third line expresses the decoupling at large time separations consistent with cluster decomposition. The factor $\expval{\d(\bco(0)-u)}{\O}$ can be calculated as $\lim_{N_t \to \ii} \Tr \lp e^{-N_t H} \d(\bco(0)-u)\rp$. Therefore, one may write
\bad 
\expval{\d(\bco(0)-u)}{\O} = \lim_{N_t \to\ii }\oint_{\phi(\b)=\phi(0)} [\cald \phi] e^{-S[\phi]}\d(\bco(0)-u).
\label{eq:delta_expval}
\ead
To evaluate this,  $N$ $d$-dimensional spatial boxes $B_1,\cds,B_N$ centered at an equispaced set of points $\{\vec x_i\}$ are defined, each encompassing $\ff{V}{a^d N}$ sites. Coarse-grained quantities $\bco_i(t)$ are defined as:
\bad
\bco_i(t) = \ss{\ff{Na^d}{V}}\sum_{\vec x \in B_i}\co(t,\vec x),
\ead
such that
\bad 
\bco(t) = \ff{1}{\ss N}\sum_{i=1}^N \bco_i(t).
\ead 

By integrating over all remaining variables, Eq. \eqref{eq:delta_expval} can be expressed as:
\bad 
\expval{\d(\bco(0)-u)}{\O} &= \int \lk \prod_{i=1}^N du_i \rk P_{\bco_1,\cds,\bco_N}(u_1,\cds,u_N) \\ 
&\quad \quad \times \d\lp \ff{u_1+\cds+u_N}{\ss N}-u\rp,
\ead 
where here and henceforth the abbreviation $\bco_i \equiv \bco_i(0)$ will be used and $P_{\bco_1,\cds,\bco_N}(u_1,\cds,u_N)$ is the joint probability density of  events in which  each $\bar{\cal O}_i$ takes the value $u_i$ and is normalised such that 
\bad
\int \lk \prod_{i=1}^{N} du_i \rk P_{\bco_1,\cds,\bco_N}(u_1,\cds,u_N) = 1,
\ead

If the box sizes are larger than  a few correlation lengths, the $\bco_i$ are independent of each other up to exponentially small effects proportional to $e^{-ml}$, where $m$ is the mass gap specified after Eq.~\eqref{eq:gapstate} and $l$ is the distance between centers of  neighbouring boxes. Consequently, $P_{\bco_1,\cds,\bco_N}(u_1,\cds,u_N) \approx \prod_{i=1}^{N} P_{\bco_i}(u_i)$. Such an approximation becomes exact in the infinite volume limit, and with $N$ and $l$ both taken to infinity\footnote{The following set of manipulations follows results of Andrey Andreyevich Markov, see Ref.\ \cite{Kac} for further details.}:
\begin{widetext}
\bad 
\lim_{V\to \ii} \expval{\d(\bco-u)}{\O} &= \lim_{N,l\to \ii} \int \lk \prod_{i=1}^N du_i \rk P_{\bco_1,\cds,\bco_N}(u_1,\cds,u_N)  \d\lp \ff{u_1+\cds+u_N}{\ss N}-u\rp \\  &=\lim_{N\to \ii} \int \lk \prod_{i=1}^N du_i  P_{\bco_i}(u_i) \rk \d\lp \ff{u_1+\cds+u_N}{\ss N}-u\rp \\
&= \lim_{N\to\ii} \ff{1}{(2\pi)^{N+1}}\int_{-\ii}^\ii d\l \int \lk \prod_{i=1}^N du_i dk_i \ti P_{\bco_i}(k_i)\rk  e^{i\lp k_1u_1 + \cds + k_N u_N \rp} e^{i\l\lp u - \ff{u_1+\cds+u_N}{\ss N}\rp}\\
&=\lim_{N\to\ii}\ff{1}{2\pi}\int_{-\ii}^\ii d\l e^{i\l u} \lk \ti P_{\bco_1}\lp \ff{\l}{\ss N}\rp \rk^N, \\
\ead 
\end{widetext}
where $\ti P_{\bb \co_i}(k_i)$ is the Fourier transform of $P_{\bb \co_i}(u_i)$. In the fourth line, that fact that $\ti P_{\bb \co_i}(k) = \ti P_{\bb \co_1}(k)$ by  translational invariance has been used. Note that  $\ti P_{\bco_1}(0) = 1$ as $P_{\bco_1}(\cd)$ is a normalized probability density function.  Similarly, $\ti P_{\bco_1}'(0) = 0$ as $\ev{\bco_1} = 0$ by assumption, and setting $\ti P_{\bco_1}''(0) = -\s_{\bco_1}^2$, $\ti P_{\bco_1}(\l)$ can be expressed as:
\bad
\ti P_{\bco_1}(\l) &= e^{-\ff 1 2 \s^2_{\bco_1}\l^2 + \co \lp \l^3 \rp }. 
\\ \nonumber
\ead  
This implies that as $N \to \ii$:
\bad 
\lim_{N\to \ii} \lk \ti P_{\bco_1}\lp \ff{\l}{\ss N} \rp \rk^N = e^{-\ff{1}{2} \s^2_{\bco_1} \l^2}, \\ \nonumber
\ead 
and by integrating over $\l$ it is clear that:
\be 
\lim_{V \to \ii} \expval{\d(\bco-u)}{\O} = \ff{1}{\ss{2\pi}\s_{\bco_1}} e^{-\ff{u^2}{2\s_{\bco_1}^2}}.
\ee
\nl
Therefore in the infinite-volume limit, the joint probability density $P_{\bco(t),\bco(0)}(u,v)$ is given as:
\be 
\lim_{V\to \ii} P_{\bco(t),\bco(0)}(u,v) = \ff{1}{2\pi\s^2_{\bco_1}} e^{-\ff{u^2+v^2}{2\s_{\bco_1}^2}} + \co\lp e^{-m \min{(t,\b-t)}} \rp. 
\ee

As a consequence, the distribution of the product  $\bco(t)\bco(0)$  in the same limit is:
\begin{widetext}
\bad 
\lim_{V\to\ii} P_{\bco(t)\bco(0)}(q) &= \int du dv \lk \lim_{V\to \ii}P_{\bco(t),\bco(0)}(u,v)\rk \d(q-uv) \\ 
&\stackrel{ t,N_t \to \ii}{\longrightarrow} \ff{1}{2\pi\s_{\bco_1}^2}\int d\o \ff{e^{i\o q}}{\ss{\lp \o-\ff{i}{\s_{\bco_1}^2}\rp \lp \o+\ff{i}{\s_{\bco_1}^2}\rp}} + \co\lp e^{-m \min{(t,\b-t)}} \rp.
\ead 
\end{widetext}
This expression reduces to Eq. \eqref{eq:FreeProductPIntegral} with $\o_\pm = \ff{1}{\s_{\bco_1}^2}$, so finally from Eq. \eqref{eq:FreeProductP} one obtains\footnote{It must be stressed that the limit $N_t \to \ii$ must be taken before the limit $t \to \ii$ while the limit $V \to \ii$ can be interchanged with the limits $t \to \ii$ and $N_t \to \ii$.}:
\be 
\lim_{t \to \ii} \lim_{N_t \to\ii}\lk \lim_{V\to \ii} P_{\bco(t)\bco(0)}(q)\rk = \ff{1}{\pi \s_{\bco_1}^2}K_0\lp \ff{\abs{q}}{\s_{\bco_1}^2}\rp.
\label{eq:generic}
\ee 

\begin{figure}[!b]
    \centering
    \includegraphics[scale=0.35]{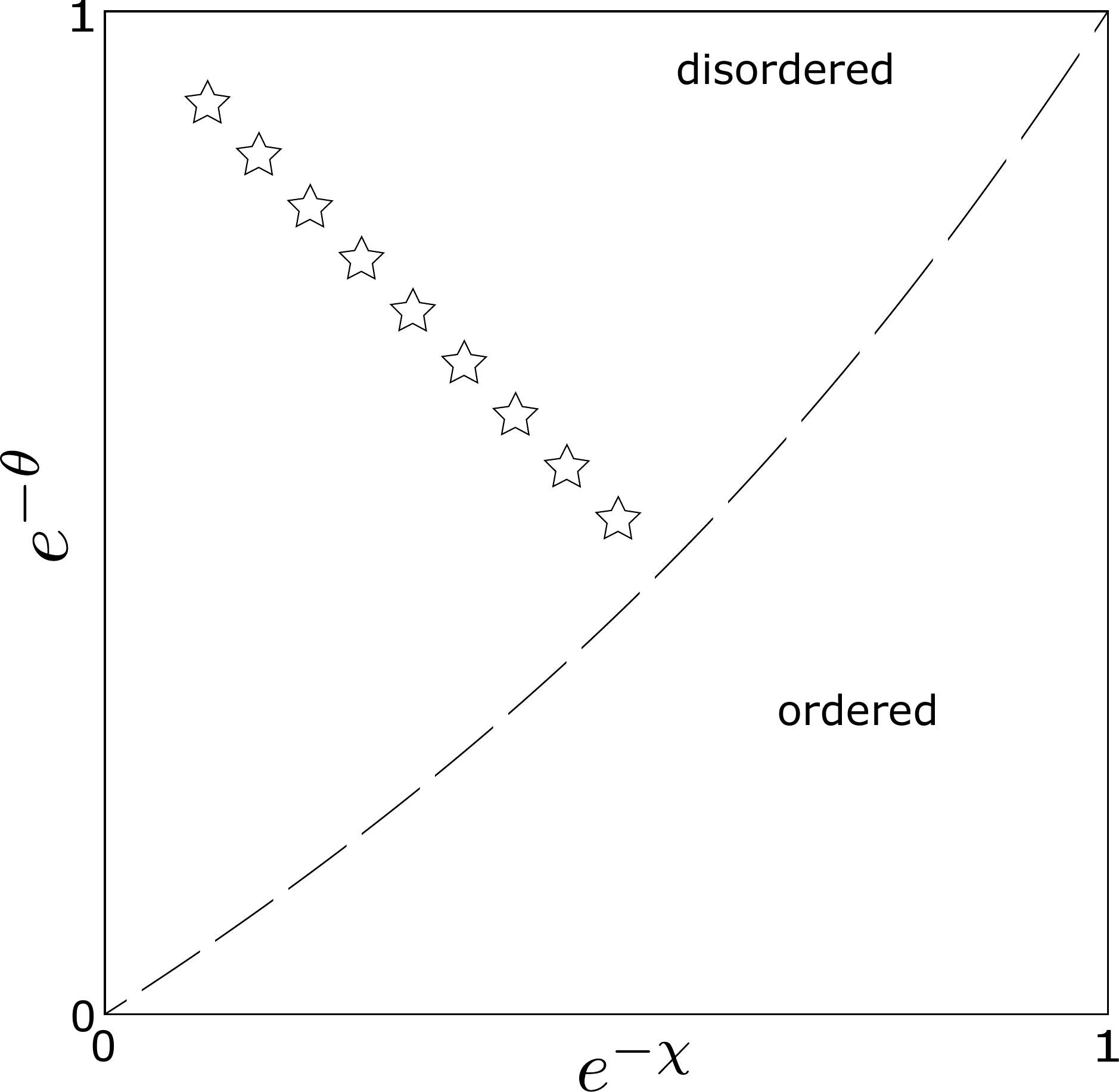}
    \caption{Schematic phase diagram corresponding to the action given in Eq. \eqref{eq:Sthetachi}. The dashed line separates the disordered and ordered phases and is known to be a second order transition. The stars correspond to the values of the couplings at which simulations are performed. A numerical determination  of this phase diagram is given by Fig. 8 of Ref.~\cite{PhysRevB.42.2445}.} 
    \label{fig:Sthetachi}
\end{figure}

To test the validity of Eq. \eqref{eq:generic}, interacting $\phi^4$ theory in two dimensions is investigated numerically following Ref. \cite{PhysRevB.42.2445}. The action for this theory is:
\be 
S = \sum_{i}\lk -\ff b 2 \phi_i^2 + \ff u 4 \phi^4_i + \ff{K}{2} \sum_{\hat \mu = 1}^2 \lp \phi_{i+\hat \mu} - \phi_i\rp^2\rk,
\ee 
where $i$ labels the sites, $b,u,K$ are couplings and $\hat \m$ labels the directions. Through the rescalings 
\bad
\phi_i &= \ff{1}{\ss K}\varphi_i, \\ 
b &= \theta K, \\
u &= \chi K^2,
\ead
the action be rewritten as 
\be 
S = \sum_i \lk \lp 2-\ff \theta 2\rp \varphi_i^2 + \ff \chi 4 \varphi_i^4 \rk - \sum_{\lc i j \rc} \varphi_i \varphi_j, 
\label{eq:Sthetachi}
\ee 
where $\sum_{\lc i j \rc}$ indicates summation over all pairs of neighbouring points. A schematic illustration of the phase diagram corresponding to the above action is given in the Fig.\ \ref{fig:Sthetachi} in terms of exponentials of the couplings $\theta$ and $\chi$.

Defining the one-parameter path through the coupling space
\bad 
e^{-\chi} &= s, \\
e^{-\theta} &= 1-s,
\label{eq:sdef}
\ead 
calculations have been performed  for a lattice size of $N_s\times N_t=40\times 40$ and sample size ${\cal N}=10^6$ for $s = 0.05 \times k$ where $k \in \{ 2,\cds,10\}$ using a publicly available code \cite{agimenezromero}. For these values of $s$, the system is in the disordered phase, and larger $s$ values correspond to smaller renormalised mass values, $m$,  closer to the critical line.
The parameter $\s^2_{\bco}$ in Eq. \eqref{eq:generic} is determined through Eq. \eqref{eq:revtrans}. To quantify how well Eq. \eqref{eq:generic} describes the numerical data, the total variance between the empirical distribution $E_t(q)$ at time $t$ and the expected asymptotic distribution of Eq. \eqref{eq:generic} is calculated:
\bad
\calt(t) \equiv \ff 1 2 \int dq\, \abs{E_t(q)-\ff{1}{\pi \s_{\bco_1}^2}K_0\lp \ff{\abs{q}}{\s_{\bco_1}^2}\rp}.
\ead 

\begin{figure}[!t]
    \centering
    \includegraphics[scale=0.55]{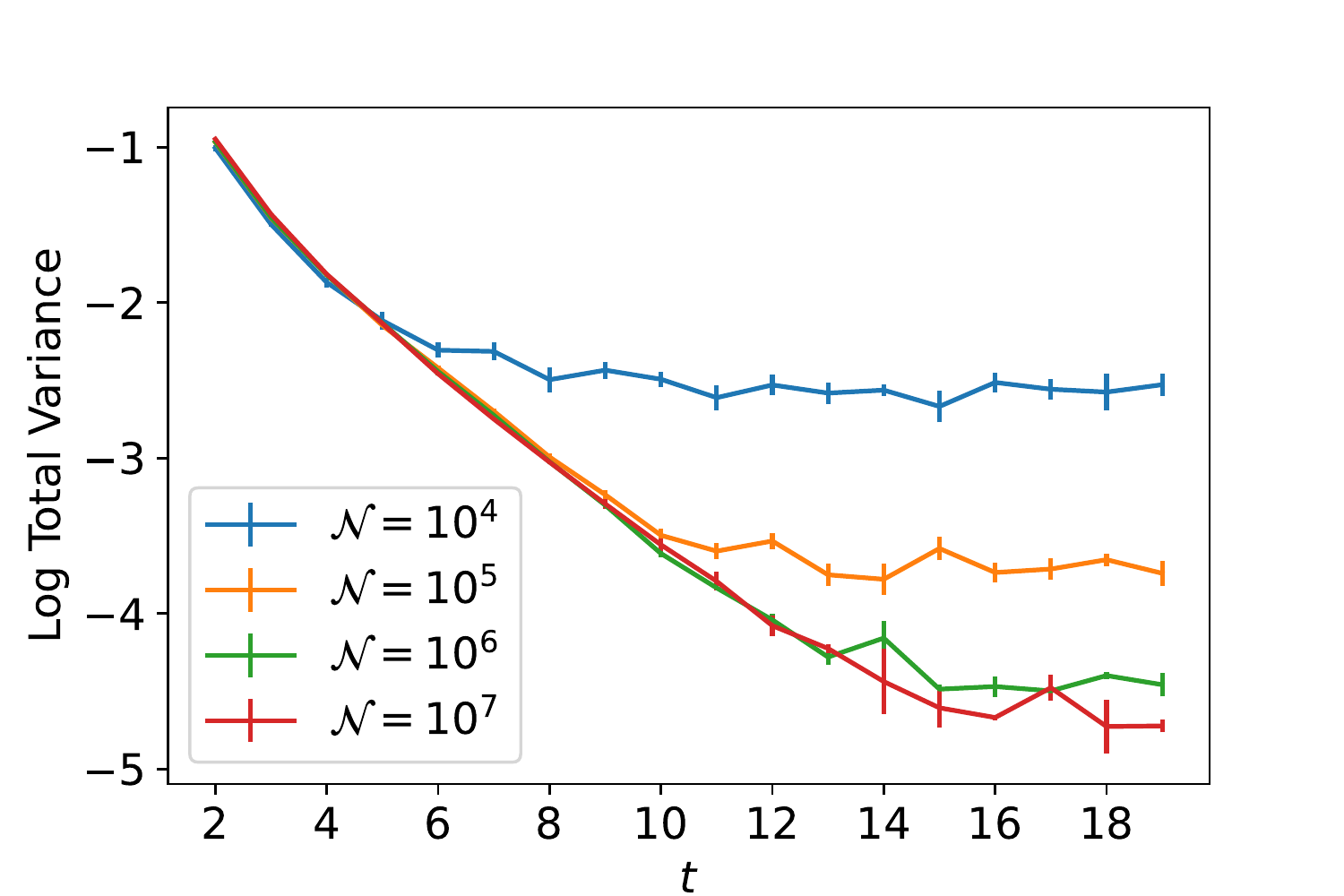}
    \caption{The logarithm of the total variance as a function of time separation for the $\phi^4$ theory in the disordered phase. Calculations are performed for $N_s \times N_t = 40 \times 40$, with $s=0.5$, and results for two different sample sizes ${\cal N}=\{10^4,10^5, 10^6, 10^{7}\}$ are shown.}
    \label{fig:TVfit_s=0.5}
\end{figure}

It is expected that the total variance vanishes in the large $t,\b$ limit. The logarithm of the total variance versus $t$ is shown in Fig.\ \ref{fig:TVfit} where it is seen that the total variance decreases as $t$ is increased until it reaches a plateau value that appears to be independent of $s$. Note that, as is seen in Eq. \eqref{eq:generic}, $\ff{1}{\pi \s_{\bco_1}^2}K_0\lp \ff{\abs{q}}{\s_{\bco_1}^2}\rp$ is the dominant term in the expansion of $P_{\bco(t)\bco(0)}(q)$ at large times and arises from the contributions of  the vacuum state. The subleading terms in this expansion are due to excited states with vanishing spatial momentum.\footnote{Since $\bco(0)$ is invariant under spatial translations, $\d(\bco(0)-u)$ is also invariant. Therefore $\mel{\Psi}{\d(\bco(0)-u)}{\O}$ is non-vanishing only if $\ket \Psi$ has vanishing spatial momentum. This proves that $P_{\bco(t),\bco(0)}(u,v)$ can be expanded in eigenstates with vanishing spatial momentum and same is the true for $P_{\bco(t)\bco(0)}(q)$.} It follows that: 

\begin{figure}[t]
    \centering
    \includegraphics[scale=0.50]{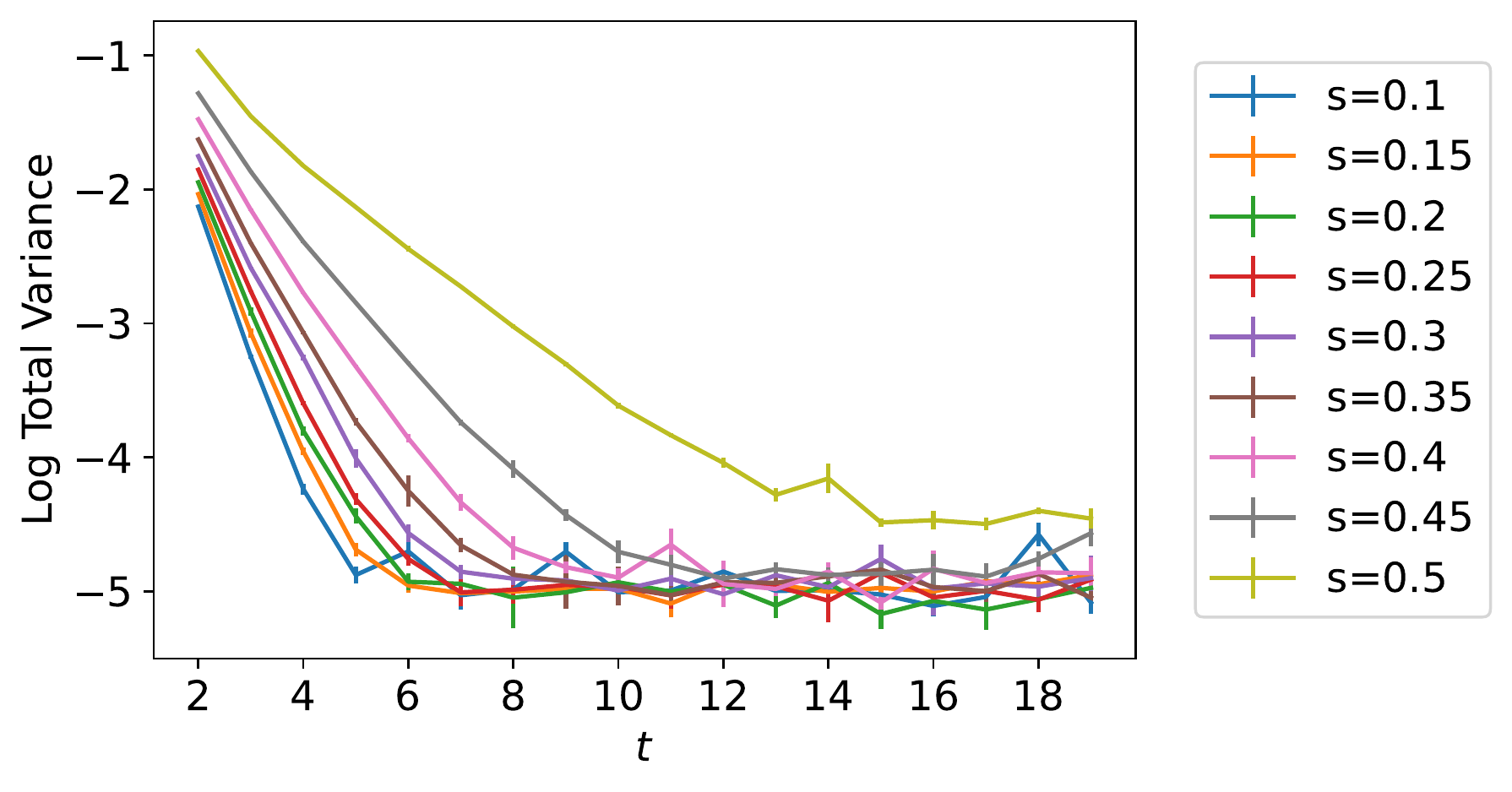}
    \caption{The logarithm of the total variance as a function of time separation for the $\phi^4$ theory in the disordered phase for  various values of the parameter $s$ in Eq. \eqref{eq:sdef}. Calculations are performed with $N_s \times N_t = 40 \times 40$, and a sample size ${\cal N}=10^6$. }
    \label{fig:TVfit}
\end{figure}
\begin{widetext}
\bad 
\lim_{V\to\ii}P_{\bco(t)\bco(0)}(q) &= \ff{1}{\pi \s_{\bco_1}^2}K_0\lp \ff{\abs{q}}{\s_{\bco_1}^2}\rp + e^{-mt} A_m(q) 
+\int_{\ti m}^\ii d\mu\, e^{-\mu t}\r(\mu) A_\mu(q),
\ead 
where $\ti m$ is the mass of the second excited state with zero spatial momentum, $\rho(\mu)$ is the density of states, and $A_m$ and $A_\mu$ are $t$-independent quantities. In this expression, it is assumed that $0 \ll  t \ll \beta/2$ so that effects of the finite temporal extent can be ignored. Considering the total variance as a function of time, it is seen that
\bad
\log(\calt_{th}(t)) &= -\log(2)-mt + \log\lp \int dq \abs{A_m(q) +  \int_{\ti m}^\ii d\mu\, e^{-(\mu-m) t}\r(\mu) A_\mu(q)}\rp
\label{eq:logt},
\ead 
\end{widetext}
where $\calt_{th}(t)$ is the infinite sample size limit of $\calt(t)$. At values of $t$ for which $A_m(q)$ dominates the remaining contributions, one expects to find linear behaviour with  slope $-m$. Such behaviour is seen in Fig.\ \ref{fig:TVfit} for some values of $t$, but at larger $t$, a constant behaviour is seen. The above expansion assumes the distribution $\lim_{V \to\ii}P_{\bco(t)\bco(0)}(q)$, while numerical calculations determine $E_t(q)$, that deviates from the exact distribution at finite statistical sampling. Since $e^{-mt}A_m(q)$ vanishes as $t$ increases, the deviation of $E_t(q)$ from the exact distribution will be larger than $e^{-mt}A_m(q)$ at large times, invalidating the above expansion. The main contribution to this deviation is expected to be due to the finite sample size, since convergence to normality is generically very robust if $m N_s \gg 1$. Additionally, since $t\sim N_t/2$ in the figure, effects of the finite temporal extent may need to be accounted for. This expectation is numerically confirmed for the system under consideration in Fig.\ \ref{fig:TVfit_s=0.5} where results of calculations are shown for $s=0.5$ for sample sizes ${\cal N}\in\{10^4,10^5,10^6,10^7\}$. 

\begin{figure}[!h]
    \centering
    \includegraphics[scale=0.55]{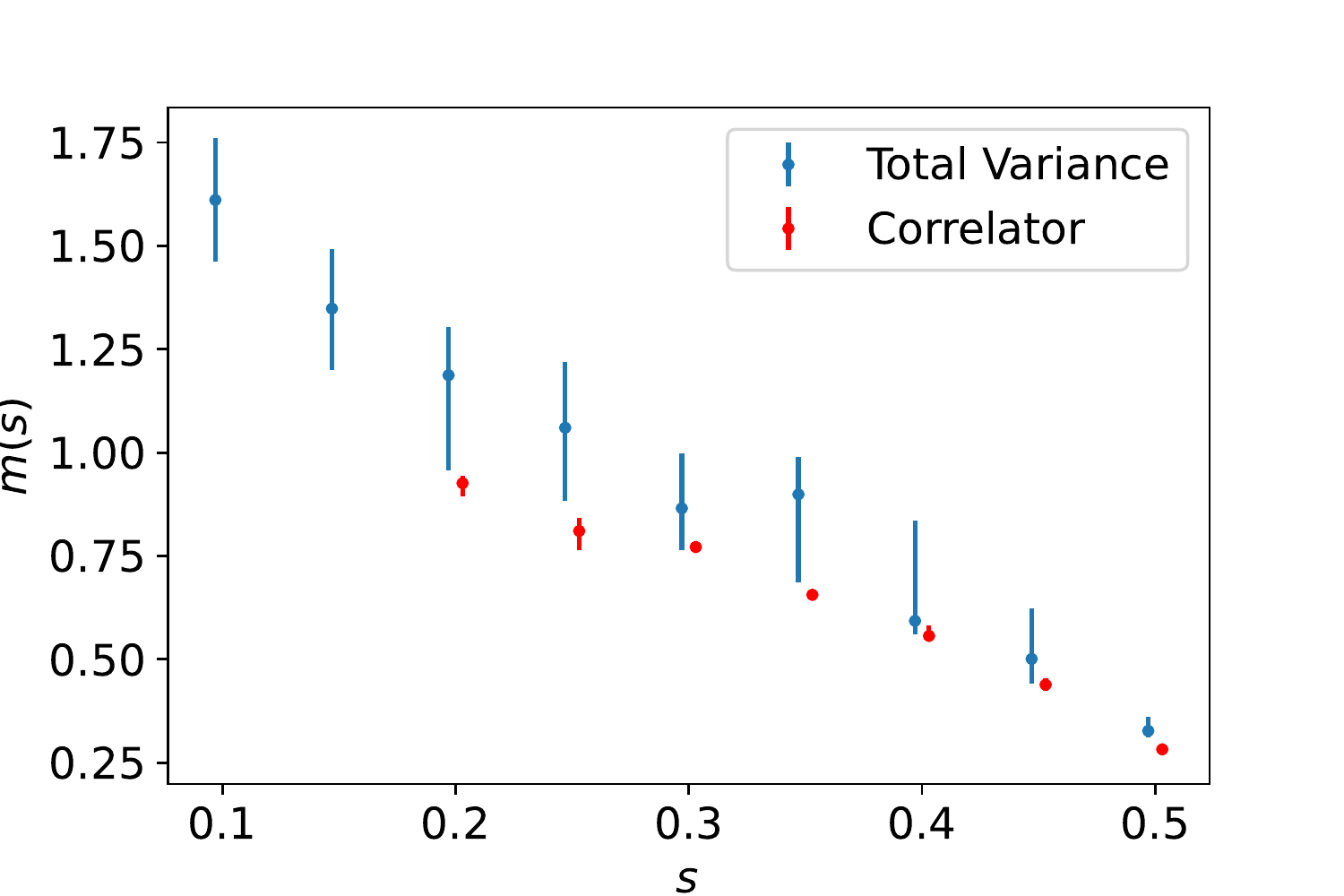}
    \caption{Estimates of $m(s)$  from the time derivative of the logarithm of total variance are shown in blue, while those from the time-dependence of the correlation function $\langle C(t) \rangle$ are shown in red.
    Error bars represent the standard deviations  calculated using 250 bootstrap resamplings.}
    \label{fig:m_estimate}
\end{figure}

Since Eq.~\eqref{eq:logt} depends on the mass of the  lowest energy state with the correct quantum numbers (Eq.~\eqref{eq:gapstate}), the time-dependence of the total variance between the empirical  distribution and the asymptotic expectation can be used to extract the corresponding mass, $m$.
Estimates of $m$  from the exponential falloffs seen in Fig. \ref{fig:TVfit} are  shown in Fig.\ \ref{fig:m_estimate} as a function of $s$. Estimates from fits to the time dependence of the correlation function $\langle C(t) \rangle$ are also shown. 
The fits to extract these masses are discussed in  Appendix \ref{app:fits}.
As can be seen in Fig.~\ref{fig:m_estimate}, the masses extracted from the total variance and the time-dependence of the correlation function itself are consistent although the total variance provides a less precise estimate for the parameters used in this study. 

\section{Summary and Outlook}

In this work, the statistical behaviour of correlation functions of bosonic lattice field theories at large Euclidean time-separations has been investigated. The exact distribution of bilocal correlation functions  in free 
scalar field theory was determined and numerical Monte-Carlo calculations were seen to converge to this distribution. It was also shown that the distributions of many correlation functions in a general class of bosonic lattice field theories approach the same universal distribution in the large-time limit. Numerical tests also confirmed this behaviour.

Extension of these results to other phenomenologically relevant theories such as lattice Quantum Chromodynamics is possible and may help in diagnosing the signal-to-noise problem that plagues calculations of many quantities. A more thorough understanding of noise in Monte-Carlo sampling of lattice field theories along directions analogous to those pursued  here, may lead to improved strategies for extracting physical information. In particular, tests of empirical distributions against the expected asymptotic distribution of correlation functions at large time may build confidence that a given level of sampling is sufficient for robust physical results to be determined. A deeper exploration of this direction is left to future work.

\begin{figure*}[!t]
\includegraphics{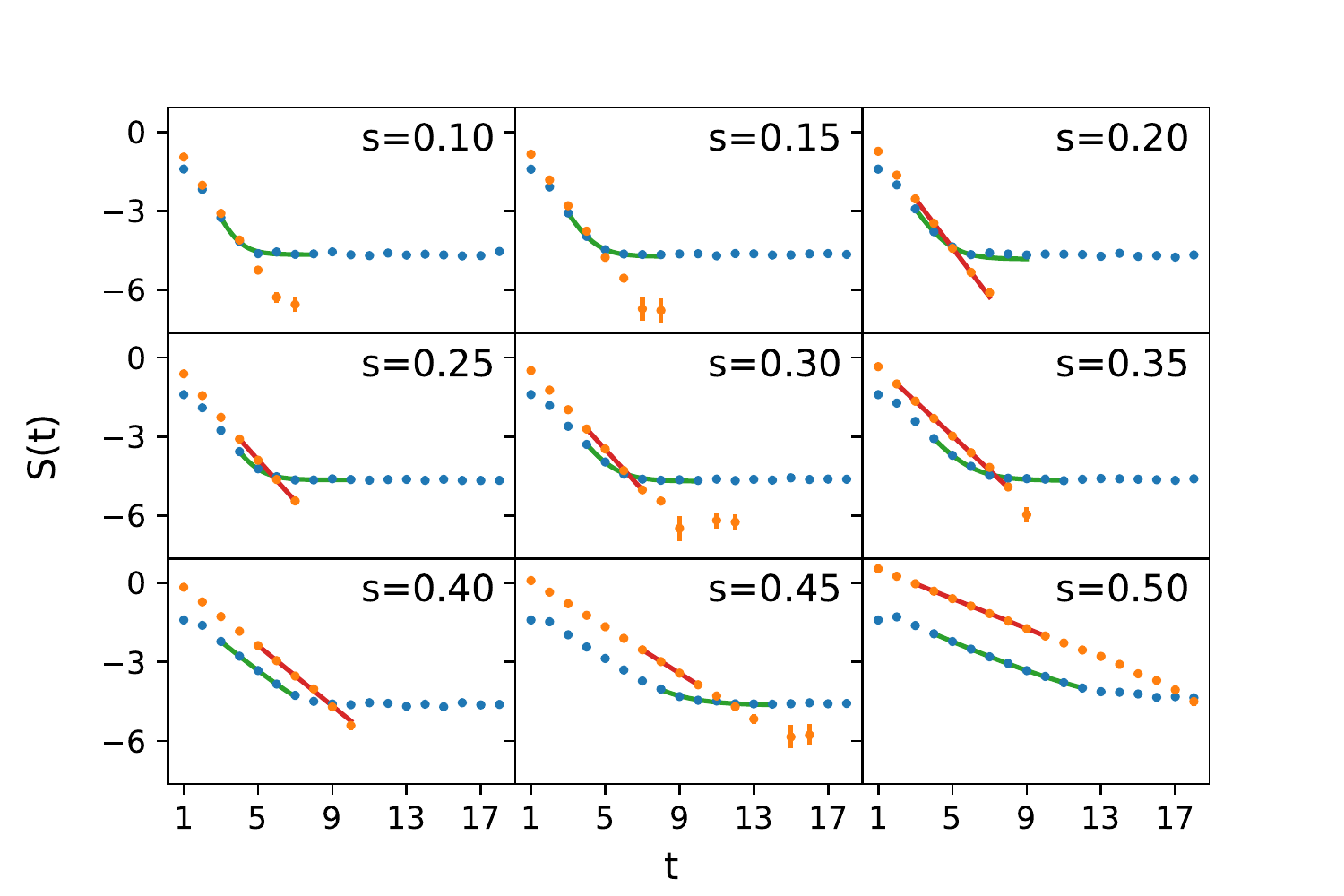}
\caption{{The behaviour of $S(t) = \log \calt(t)$ (blue) and $S(t) = \log \lc C(t)\rc$ (orange) for $t=1,\cds,18$ and $s = 0.1,0.15,\cds,0.5$ is shown. Uncertainties are determined through the bootstrap method and the best fits $\hat f$ (defined in Appendix \ref{app:fits}) are shown as smooth curves over the time range of the best fit. }}
\label{fig:combined}
\end{figure*}

\acknowledgements{
We are grateful for insightful discussions with P. Shanahan and M. L. Wagman.
This work is supported by the National Science Foundation under Cooperative Agreement PHY-2019786 (The NSF AI Institute for Artificial Intelligence and Fundamental Interactions, http://iaifi.org/) and by the U.S.~Department of Energy, Office of Science, Office of Nuclear Physics under grant Contract Number DE-SC0011090. WD is also supported by the SciDAC4 award DE-SC0018121. 
} 

\appendix
\section{Fits to correlation functions and total variance}
\label{app:fits}

The fits to extract masses from the correlation functions and total variance functions are performed as follows. For each value of $s$, $B=250$ bootstrap samples are generated, each having $10^6$ samples. For each bootstrap sample, $\calt(t)$ and $C(t) \equiv \bco(t)\bco(0)$ are calculated. 
In order to determine the mass, fits of the form $f_{\cal T}(t)=A+Be^{-E_{\calt}t}$ must be performed to $\calt(t)$ and fits of the form $f_C(t)=De^{-E_{C}t}$ must be performed to $C(t)$. In what follows, only the extracted energies $E_C$ and $E_{\cal T}$ are  meaningful and referred to generically as $E$. For each $s$, bounds of acceptable fit ranges $t^\calt_<, t^\calt_>, t^C_<, t^C_>$ are set as follows: $t^C_<$ is always chosen to be $t^C_< =1$; $t^\calt_<$ is chosen to be minimum $t$ that satisfies $\log \calt(t') < \log \calt(t)$ for all $t' > t$; $t^C_>$ is chosen to be minimum value of $t$ that satisfies $\ev{C(t)}<0$; $t^{\calt}_>$ is chosen to be $t$ value where $\log \calt(t)$ becomes consistent with a constant. 
Fits to the two quantities are performed over subranges   $(t_{\min},t_{\max})$ within these bounds that satisfy the inequalities $t_< \leq t_{\min} \leq \max\lp2,t_{\max}-3\rp$ and $5 \leq t_{\max} \leq \max\lp t_{>}, 7 \rp$. Fully correlated fits are performed for both $\calt(t)$ and $C(t)$ with covariance matrices calculated through optimal shrinkage \cite{shrinkage}. Only the fits that satisfy $\ff{\chi^2}{d.o.f.}<1.5$ are deemed acceptable and kept for further analysis. These surviving fits are labelled by $f =\in \{ 1,\cds, F\}$ and are assigned a weight $W_f \propto \ff{1}{\d E_f^2}e^{-\ff{\chi^2}{d.o.f.}}$ where $\d E_f^2$ is a measure of the statistical variance of a given fit. Defining the  deviations $\Delta E_f^b = E_f^b-E_f$ on each bootstrap $b$ for each fit range $f$ ($E_f$ is the bootstrap mean for fit range $f$), this statistical uncertainty is defined by
\bad 
\d E_f = \ff 1 2 \lk Q_{\ff 5 6}\lp \{\D E_f\} \rp - Q_{\ff 1 6}\lp \{\D E_f\}\rp \rk
\ead 
where $Q_q\lp \cd \rp$ is the quantile function, whose argument is a bootstrap set. A  68\% confidence interval is given as 
$\lp Q_{\ff{1}{6}}\lp \{E\} \rp - \d E_{\hat f},Q_{\ff{5}{6}}\lp \{E\} \rp + \d E_{\hat f} \rp$ where $\{E\}= \{E_f \rvert f=1,\cds,F\}$ and $\hat f = \argmax_f W_f$. In Fig. \ref{fig:combined}, the calculated values of $\log\langle C(t)\rangle(t)$ and ${\cal T}(t)$ are shown as a function of $t$ for each  $s \in \{0.10,0.15,\cds,0.50\}$. 
The resulting masses extracted from the two data sets are shown in Fig. \ref{fig:m_estimate} as a function of $s$. 

\bibliography{refs}

\end{document}